# Growth of Large-Area Graphene Films from Metal-Carbon Melts

Shaahin Amini[1*], Javier Garay[1], Guanxiong Liu[2], Alexander A. Balandin[2], Reza Abbaschian[1]

[1]Department of Mechanical Engineering,
[2]Department of Electrical Engineering,
University of California Riverside, Riverside, CA 92507 USA

[*] To whom correspondence should be addressed, E-mail: samini@engr.ucr.edu

**Abstract**

We have demonstrated a new method for the large-area graphene growth, which can lead to a scalable low-cost high-throughput production technology. The method is based on growing single-layer or few-layer graphene films from a molten phase. The process involves dissolving carbon inside a molten metal at a specified temperature and then allowing the dissolved carbon to nucleate and grow on top of the melt at a lower temperature. The examined metals for the metal – carbon melts included copper and nickel. For the latter, pristine single layer graphene was grown successfully. The resulting graphene layers were subjected to detailed microscopic and Raman spectroscopic characterization. The deconvolution of the Raman 2D band was used to accurately determine the number of atomic planes in the resulting graphene layers and access their quality. The results indicate that our technology can provide bulk graphite films, few-layer graphene as well as high-quality single layer graphene on metals. Our approach can also be used for producing graphene-metal thermal interface materials for thermal management applications.

## 1-Introduction

Graphene is a two dimensional sheet of $sp^2$ bonded carbon atoms in a honeycomb network. This honeycomb network could be the basic building block of other important allotropes of carbonic materials: graphite, nanotubes and fullerenes[1]. Recent investigations have revealed that graphene has several unique properties including the quantum Hall effect at room temperature[2-5], ambipolar field effect[6], optical properties[7], high electron mobility[8-10] and detection of single molecule adsorption events[11]. The exceptional properties of graphene also favor its implementation in a myriad of devices. From the practical point of view, some of the most interesting graphene properties are high room temperature (RT) carrier mobility[1,4,6,12], up to ~27000 $cm^2V^{-1}s^{-1}$ and recently discovered high thermal conductivity[13-14] exceeding ~3080 W/mK. The enhanced current and heat conduction properties are beneficial for electronic, interconnect and thermal management applications of graphene. It has also been demonstrated that graphene devices can operate at very low-levels of the electronic flicker noise, which is important for applications in sensors and communications[15-17].

For graphene to be commercially implemented in devices, however, a reliable, scalable and economical processing technique must be developed. Three major requirements for suitable techniques are: (1) The technique must produce high quality crystal in a 2D lattice to ensure high mobility of carriers. (2) If the technique deals with growing of the single layer, it must provide fine control over crystallite thickness so that in the electronic device application it delivers uniform performance. (3) The process should be scalable industrially. Among several techniques[18-19], the most successful method that has resulted in isolation of single layer graphene (SLG) is the Mechanical Exfoliation (Micro mechanical cleavage) utilizing a cellophane tape[6]. Beside Mechanical Exfoliation, other promising techniques such as Epitaxial Growth[20-22], Chemical Vapor Deposition (CVD)[23-25],

Chemically Derived Graphene from Graphite Oxide[26-27] and high pressure – high temperature (HPHT) growth[28] have also been introduced.

Despite the promise of above-mentioned techniques, a method is yet to emerge as a commercially viable. The techniques thus far all have drawbacks, the largest of which are the time and equipment cost. In this letter we report the development of a very low cost yet scalable process which produces high quality graphene. The process involves dissolution of carbon atoms in a molten metal, followed by cooling the melt to allow the dissolved atoms to precipitate on top of the melt as SLG.

**2- Experimental Procedures**

The general idea in our process is to dissolve carbon atoms inside a transition metal melt at a certain temperature, and then allowing the dissolved carbon to precipitate out at lower temperatures as SLG. The schematic of the process for nickel has been shown in Fig. 1. In Fig.1 (a) nickel is melted in contact with a carbon source. This source could be the graphite crucible inside which the melting process is carried out or it could be the graphite powder or chunk sources which are simply placed in contact with the melt. Keeping the melt in contact with carbon source at a given temperature will give rise to dissolution and saturation of carbon atoms in the melt based on the binary phase diagram of metal-carbon (Fig.1 (b)). Upon lowering the temperature, solubility of the carbon in the molten metal will decrease and the excess amount of carbon will precipitate on top of the melt (Fig.1 (c)). The temperature-time diagram of the process has been shown in Fig.1 (d). The floating layer can be either skimmed or allowed to freeze for removal afterwards.

The abovementioned processing technique was utilized for copper and nickel, for which the related phase diagrams are shown in Fig.2. The processing temperatures indicated in Fig.1 has been marked in Fig.2(b). Arc melting process and melting in resistance furnace were selected for dissolving process. For the former technique a Direct-Current Electrode Negative (DCEN) process was used. The processing chamber was vacuumed and then backfilled with argon for two times. The current was chosen as 75 amps and the melting process was carried out for 20 seconds. In resistance furnace technique, the furnace was first vacuumed to $10^{-6}$ torr and then backfilled with purified argon. After reaching 1500 °C, the samples were kept for 16 hours and then cooled to ambient temperature. Both heating and cooling rates were 10 ºC/min. For Cu-C system, the melting process was carried out in graphite crucibles as carbon source. In Ni-C system a hypereutectic composition of Ni+2.35 wt% C was selected and the specified amount of carbon was added to the molten metal in the form of chunk graphite.

The samples were then investigated with optical microscope, Scanning Electron Microscope and Raman Spectroscopy. Additional investigation was done by dissolving the metal substrate away and transferring the graphene layers to a silicon wafer by the method previously reported for the transfer of carbon nanotubes and graphene layers[23,29]. To achieve this, a layer of poly methyl methacrylate (PMMA) was spin coated on the substrate (1800 rpm for 30 seconds). Afterwards, the metal substrate was etched away by a nitric acid solution (1:2) allowing the PMMA/carbonic layer to float on top of the solution. The layer was then placed on a glass substrate and washed with isopropanol and water. The dual layer of PMMA and carbonic layer was then transferred to a Si/SiO$_2$ wafer. The film was annealed at 60°C for 1 hour to adhere firmly to the target substrate. The PMMA is then dissolved with the acetone drops gradually and the carbonic layers are left on the substrate. The wafer with the carbonic layers is washed with isopropanol and dried with the nitrogen gas. Raman spectroscopy has been carried out using a Horiba Jobon Yvon micro-Raman spectrometer. All spectra were excited with visible (632.8 nm) laser light (power 3.6mW) and collected in the backscattering configuration.

The spectra were recorded with a 1800 lines/mm grating. A 100X objective to focus the excitation laser light on different spots of the samples were used.

Raman spectroscopy has been utilized as a convenient technique for identifying and counting graphene layers[30-33]. The most prominent features in the Raman spectra of graphitic materials are the G band (~1582 cm$^{-1}$), D band (~1350 cm$^{-1}$), D´ band (~1620 cm$^{-1}$) and the 2D band (~2700 cm$^{-1}$) [30,34]. The G band is Raman active for sp$^2$ carbon networks. In contrast, sp$^3$ and sp carbon show characteristic Raman features at 1333 cm$^{-1}$ (diamond) and in the range 1850–2100 cm$^{-1}$ (linear carbon chains), respectively. The D and D´ bands are defect induced Raman features. Thus these bands cannot be seen for highly crystalline graphite without any defect. The integrated intensity ratio for the D band and G band ($I_D/I_G$) is widely used for characterizing the defect quantity in graphitic materials. The 2D (or G´) band corresponds to the overtone of the D band observed in all kinds of graphitic materials and exhibit a strong Raman band which appears in the range 2500–2800 cm$^{-1}$. It has been shown[30] that the evolution of the 2D band Raman signatures with the addition of each extra layer of graphene can be used to accurately count the number of layers. A rough estimate on the number of layers can also be obtained from analysis of $I_G/I_{2D}$ ratio[30,35]. What is also important in our case is that the micro-Raman spectroscopy based graphene identification was shown to be reliable for graphene on various substrates (not only on Si/SiO$_2$)[36-37]. It also has been shown that among the metallic substrates, nickel is an appropriate one for direct Raman Spectroscopy investigation[24].

**3-Results and Discussion**

The aim of this process is to grow graphene layers as thin as a SLG. A calculation based on the lever rule[38] on Cu-C phase diagram will show that cooling from 1200 °C and 1800 °C to the melting point of copper (1080°C) will result in the formation of layers with thickness ranging from several nanometers to several micrometers respectively. Selecting higher alloying temperatures will lead to dissolution of more carbon atoms in the melt and consequently more amount of carbon will precipitate on the melt upon cooling. This will result in thicker graphite layer formation. Nevertheless, it is pertinent to point out that the precipitated graphite layer on the melt may not be uniform and its thickness varies from SLG to bulk graphite as described later.

Fig. 3(a) shows the graphite layer which has been formed on top of nickel. The film has a specific morphology of smooth surface areas separated from each other by out-of-plane faceted ridges. The areas which separate the flat regions are referred as wrinkles or creases which are marked with white arrows in Fig. 3(a). The typical size of the smooth surface regions was found to be about 50 µm. Fig.3(b) shows a typical Raman Spectrum of smooth areas. The spectroscopy shows an intense G band at 1583 cm$^{-1}$ as well as an asymmetric 2D band which exhibits two features; first a shoulder centered at 2651 cm$^{-1}$ and second the main peak centered at 2686 cm$^{-1}$. No D and D´ bands could be noticed in the spectrum. Similar Raman Spectroscopy was carried out in different spots and the results were identical. The Raman spectrum features shown in Fig.3(b) are similar to those of bulk crystalline graphite reported in literature[30]. For the selected Ni-C composition (hypereutectic alloy of Ni+2.35wt% C), upon cooling of the molten phase, based on the phase diagram a graphite shell will grow on top of the melt as the primary graphite. In fact the surface of the melt is a favorable site for heterogeneous nucleation and growth of graphite films.

Fig. 4(a) shows the magnified view of a flat area of another Ni+2.35 wt% C sample which is bounded by four triangular cross sections wrinkles. The structure at the joint between the creases has more complicated microstructure, but the faceted structure is still evident. Fig.4(b) shows the Raman spectrum which is carried out on a crease. The micro Raman spectroscopy shows the intense G band at 1582 cm$^{-1}$, and once more an asymmetric 2D band with a shoulder at 2650 cm$^{-1}$ and a main peak at 2684 cm$^{-1}$. No D and D´ bands could also be noticed in the spectrum. The wrinkled feature of the graphite layers is believed to be due to the accommodation of the thermal expansion coefficient difference between the metal substrate and the graphite layer[39]. After the graphite shell formation on top of the melt and conclusion of eutectic reaction, both nickel and graphite contract as the samples cools down. The thermal expansion coefficient of nickel[40] varies from 21.0 to 12.89×10$^{-6}$ K$^{-1}$ for the temperature range of 1200 to 27°C while the in-plane thermal expansion coefficient of graphite[41] changes from 1.25 to -1.25×10$^{-6}$ K$^{-1}$ for the same temperature range. This difference in thermal expansion coefficients will give rise to a larger lateral contraction of metal substrate than that of graphite film. As a result, a compressive biaxial stress[39] will develop on the graphite layer which consequently leads to the formation of triangular folds in the film. The wrinkle formation is schematically shown in Fig.5. Ab initio studies[41] as well as experimental results[42-43] indicate that bulk graphite below 400°C and SLG possess negative thermal coefficient that will even intensify the thermal coefficient expansion mismatch.

By comparing the smooth areas and wrinkles Raman features in Fig 3(b) and Fig 4(b), it is evident that the structure of wrinkles is identical to flat areas. In fact, the wrinkles are part of the graphite films which form during the cooling process and hold the same crystal structure of flat areas. The facets in wrinkles joint in Fig.4(b) demonstrates the crystalline structure of the wrinkles. It is believed that Weak van der Waals force among the graphene layers allows these layers to simply shift upward under biaxial stresses. Although the individual layers are rather stable owing to strong covalent bonds, the graphene layers can bend or fold without losing their crystal structure. The absence of D and D´ bands in these spectra also shows that the deformed layer with its creases are free of defect. It has been shown that deformation of graphene layers and forming creases are due to kinking [44] or twining [45]. The strong in-plane covalent bonds and the resilient structure of graphite are comprehended by the formation of these wrinkles.

A few layers graphene can also precipitate out from the melt. Fig. 6 shows the SEM photo of an electron-transparent graphitic layer on copper. The layer is thin enough to serve as a window for 5 keV electrons to pass along. Few layers graphene could also be formed at the edges of thick graphite. In transfer of graphitic layers to the $Si/SiO_2$ substrate it was observed that the color contrast at the edges of precipitated graphite islands is different. Thus there is possibility of finding few layers graphene at the edges. As a graphitic layer nucleates on the melt, it expands laterally and normally and the island edges could be as thin as few layers graphene. The schematic of this mechanism has been shown in Fig.5. One of the grown islands on copper and the Raman spectrum of its edge have been shown in Fig. 7. The spectrum features a symmetric 2D band and intense D and D´ bands. The $I_G/I_{2D}$ ratio is the evidence for the presence of 5-6 layers graphene[31]. The symmetric 2D band denotes the existence of turbostratic graphite (i.e. without ABAB stacking) [30].

The intense D and D´ peaks show layers with high amount of defects [34]. It is conceivable that this defect formation is due to entrapment of high temperature vacancies owing to the high cooling rate of copper. The thermal expansion coefficient mismatch between the substrate and graphite also gives rise to the formation of cracks. These defect formation mechanisms cause intense D and D' peak in Raman spectrum of Fig. 7. An area of a few layer graphene on nickel and its Raman spectrum has also been shown in Fig.8. The 2D band is being deconvoluted (Lorentzian analysis) for examining the number of layers. The Raman spectrum shows a G band at 1583 $cm^{-1}$. The 2D band deconvolution reveals two Lorentzian peaks at $2D_1$= 2688 $cm^{-1}$ and $2D_2$= 2660 $cm^{-1}$ ($\Delta\omega$=28 $cm^{-1}$). The Raman spectrum shows the presence of 4 layers graphene[36,46]. Missing D and D´ bands represents defect free few layers graphene on nickel. This is dissimilar to the few layers graphene forming on copper. The thermal diffusivity of copper (ability of a material to conduct thermal energy relative to its ability to store thermal energy) is nearly five times of nickel[47]. Higher thermal diffusivity of copper will lead to higher cooling rate and consequently more defect formation.

Interestingly, nickel is not Raman active and could be an appropriate substrate for direct characterization of graphene layers rather than transferring them to silicon wafer. The Raman spectrum of single layer graphene was detected in many spots on direct Raman characterization on top of the nickel. A pristine SLG and its Raman spectrum have been shown in Fig.9. The area of the grown SLG is larger than 125 $\mu m^2$. The Raman spectrum shows G band at 1583 $cm^{-1}$ and an asymmetric 2D band at 2660 $cm^{-1}$. Not observing the D peak proves that the formed SLG is pristine and defect free. The Full length at half maximum (FWHM) of grown SLG is 17 $cm^{-1}$ (compared to the reported value of 25 $cm^{-1}$)[30]. The $I_G/I_{2D}$ ratio is 4.53 and the deconvolution of 2D band showed the complete symmetry as it reported for SLG.

The amount of graphite forming on the melt and its characteristic will strongly depend on the amount of carbon dissolved in the melt and the solubility limit of carbon in the liquid as well as cooling conditions employed. For the present investigation the two alloy systems of Cu-C and Ni-C show extensive differences in solubility limit [48-49]. It approves that the Ni-C system is more conducive to the formation of large and defect free layers. However, additional studies are needed to specify the differences. The graphene-metal composite could be employed as Thermal Interface Materials (TIM) for heat dissipation purposes. Superconductivity of graphene along with the support of metal substrate could make the graphene-metal composite a superconductive filler for thermal transfer surfaces in electronic industry.

**Conclusion**

A new technique for growing large-area graphene was introduced. The technique involved dissolving carbon in a molten metal at a specified temperature and then alloying the dissolved carbon to nucleate and grow on top of the melt at a lower temperature. Detailed microscopy and micro-Raman spectroscopy were utilized to characterize the formed layers. Different morphology including thick graphite, few layers graphene and SLG were observed on metal substrate. The bulk graphite microstructure shows flat areas bounded by triangular cross section, faceted wrinkles duo to thermal expansion coefficient mismatch of metal substrate and graphite. Few Layers graphene was also observed in both nickel and copper substrate. The Raman spectroscopy proved that SLG larger than 125 µm$^2$ has been successfully grown on nickel substrate. The SLG Raman spectrum featured no D and D' band indicating the pristine and high quality nature of SLG. It is believed that among selected metals, nickel provides a better substrate for growing SLG. Since nickel is not Raman active, the direct Raman spectroscopy of graphene layers on top of the nickel is achievable. The graphene-metal composite could be utilized in thermal interface materials for thermal management applications.

**Acknowledgements**

The authors wish to thank Dr. Shahram Amini of the National Hypersonic Science Center, University of California Santa Barbara for his support and comments. AAB acknowledges support from DARPA – SRC Focus Center Research Program (FCRP) through its Center on Functional Engineered Nano Architectonics (FENA) and Interconnect Focus Center (IFC).

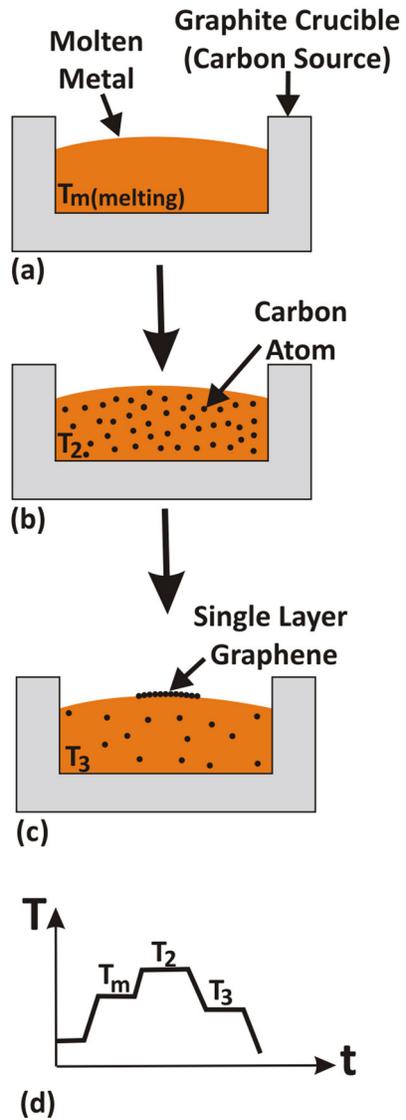

**Fig.1.** Schematic of graphene growth form molten nickel a) Melting nickel in contact with graphite as carbon source, b) dissolution of carbon inside the melt at high temperatures, and c) reducing the temperature for growth of graphene. d) shows Temperature-Time Diagram of the process

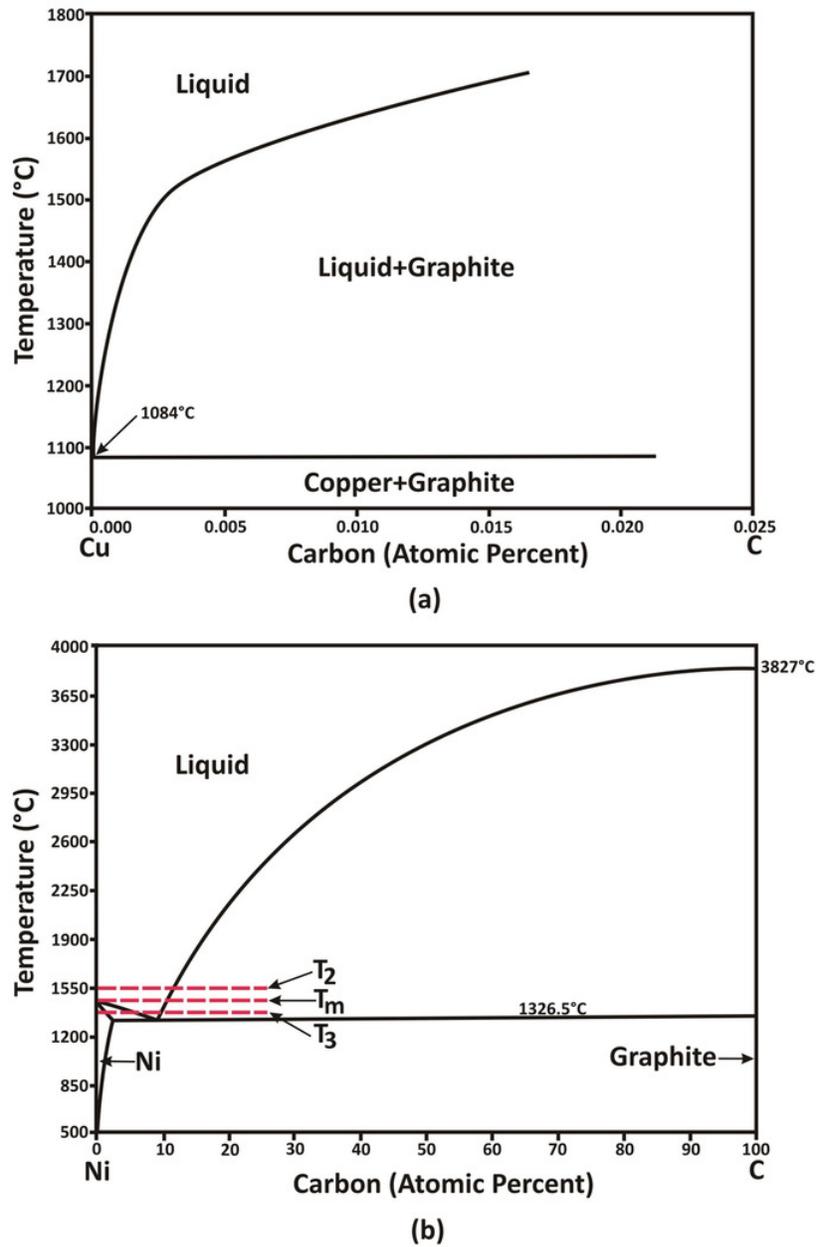

**Fig.2.** Phase diagrams of selected binary systems: Cu-C (a) and Ni-C (b), the processing temperatures of Fig.1 are marked in the Ni-C phase diagram

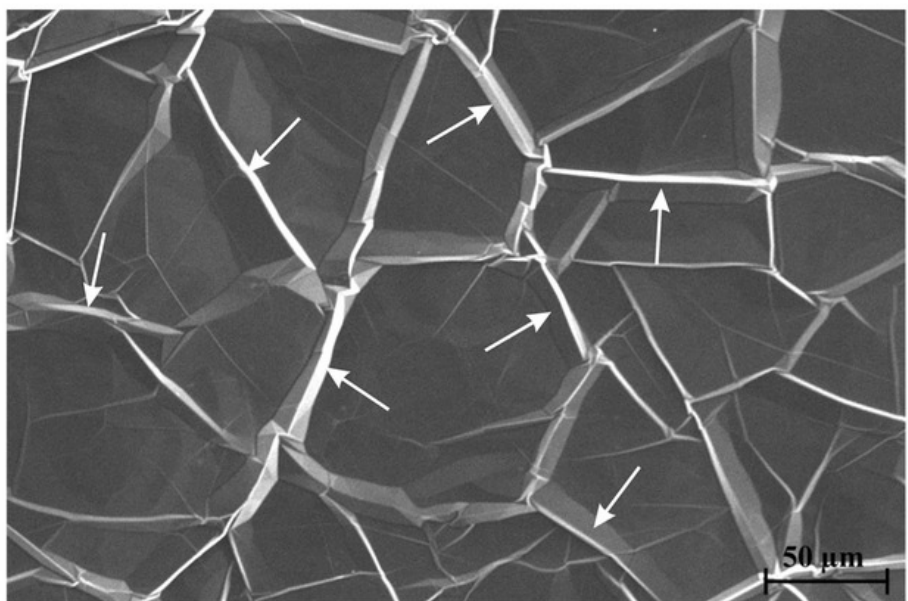

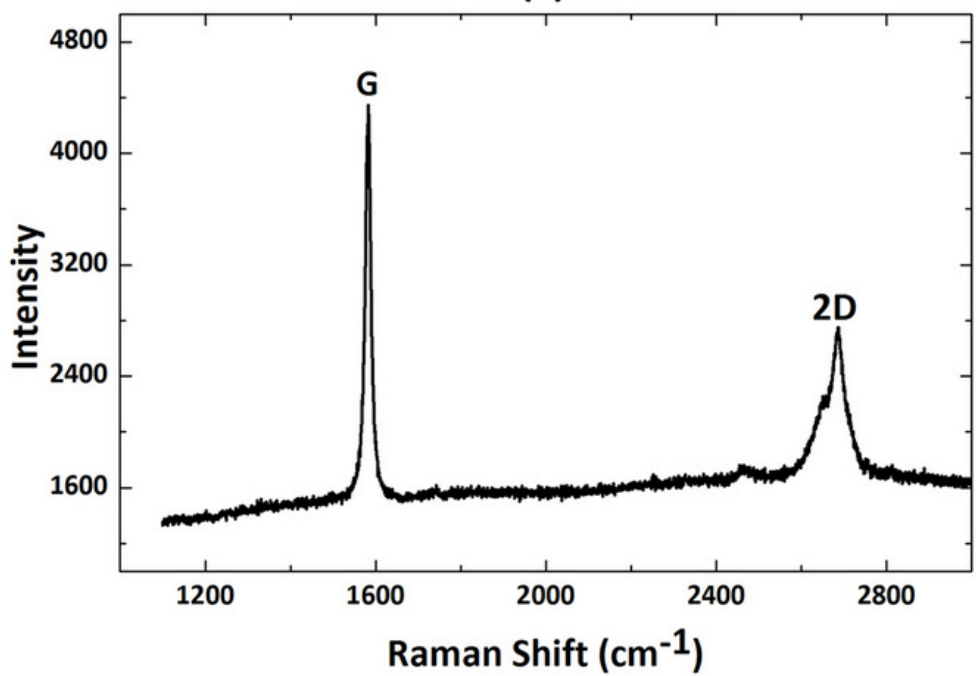

**Fig.3.** SEM Photo (Secondary Electron image) of thick graphite layer formed on top of the nickel (a) and its Raman spectrum (b), the arrows show the wrinkles

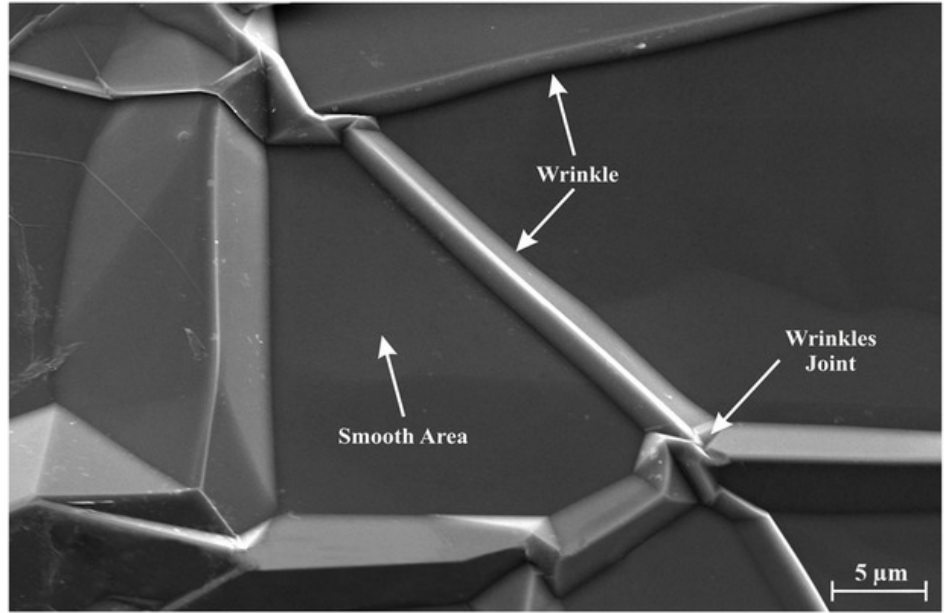

(a)

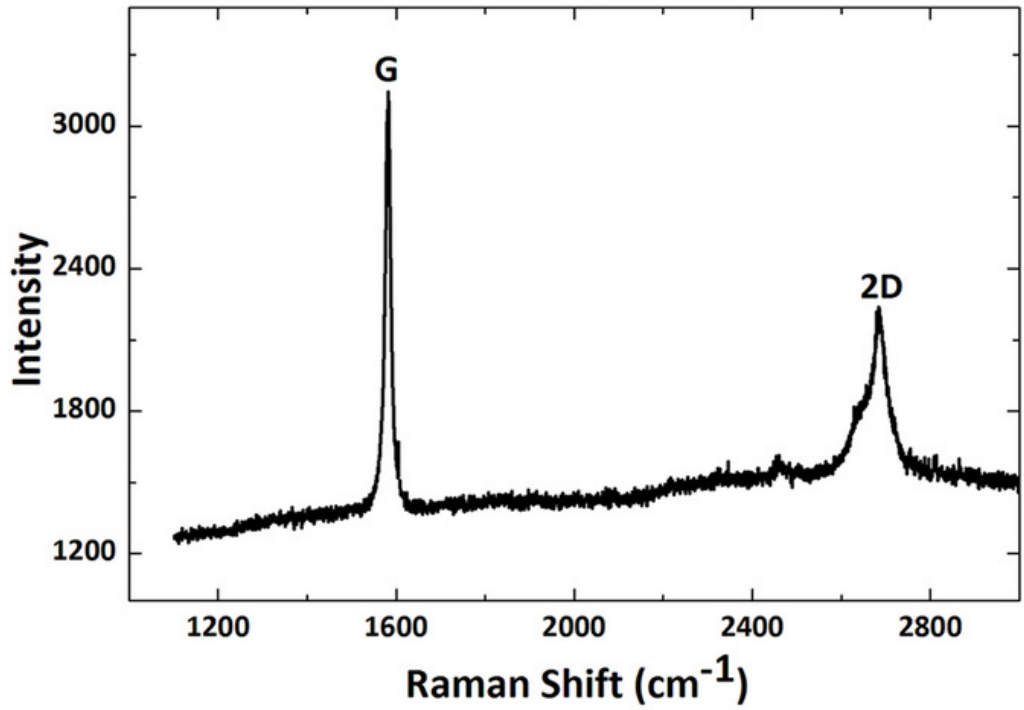

(b)

**Fig.4.** SEM Photo (Secondary Electron image) of a graphite smooth area bounded by wrinkles on top of nickel substrate (a) and the Raman spectrum of a wrinkle (b)

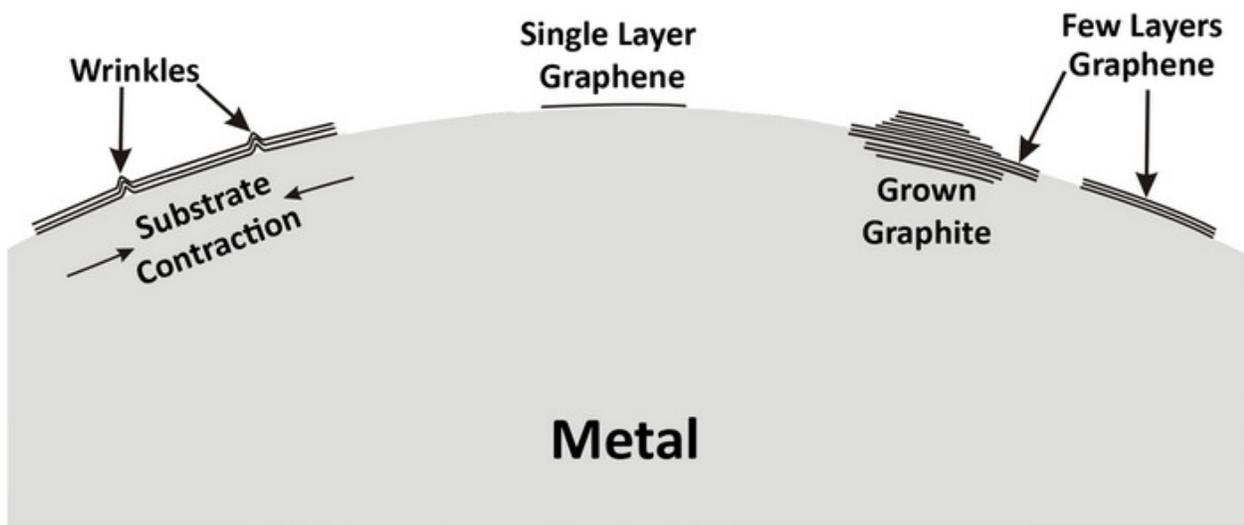

**Fig.5.** Schematic of wrinkle, single layer graphene and a few layers graphene formation

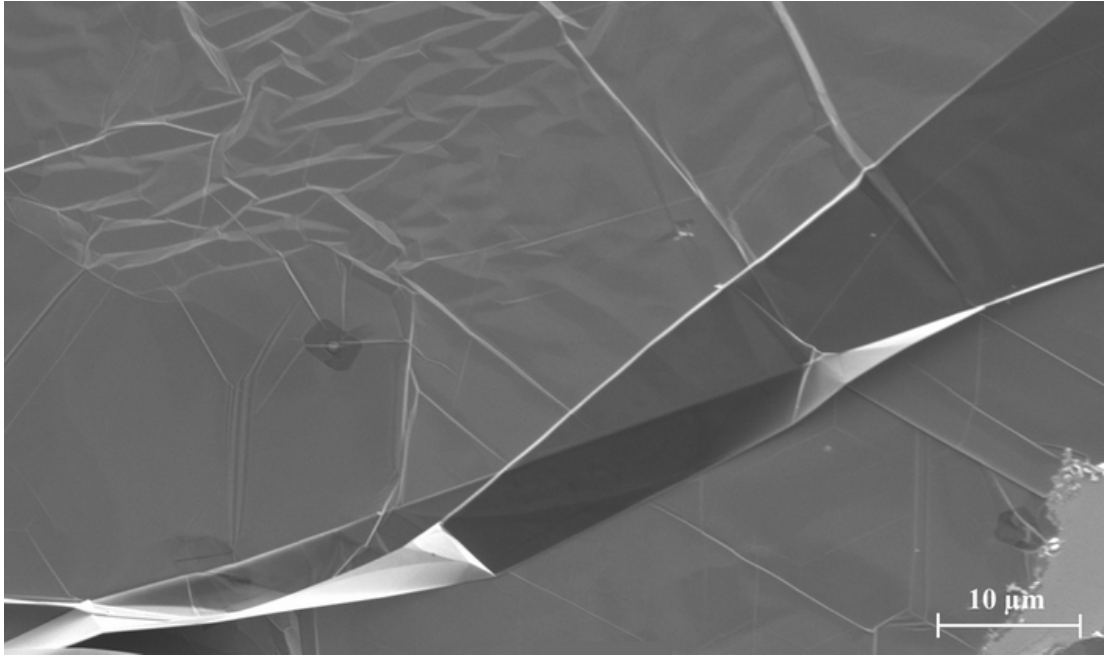

**Fig.6.** SEM photo (Secondary Electron image) of a transparent graphitic layer on nickel

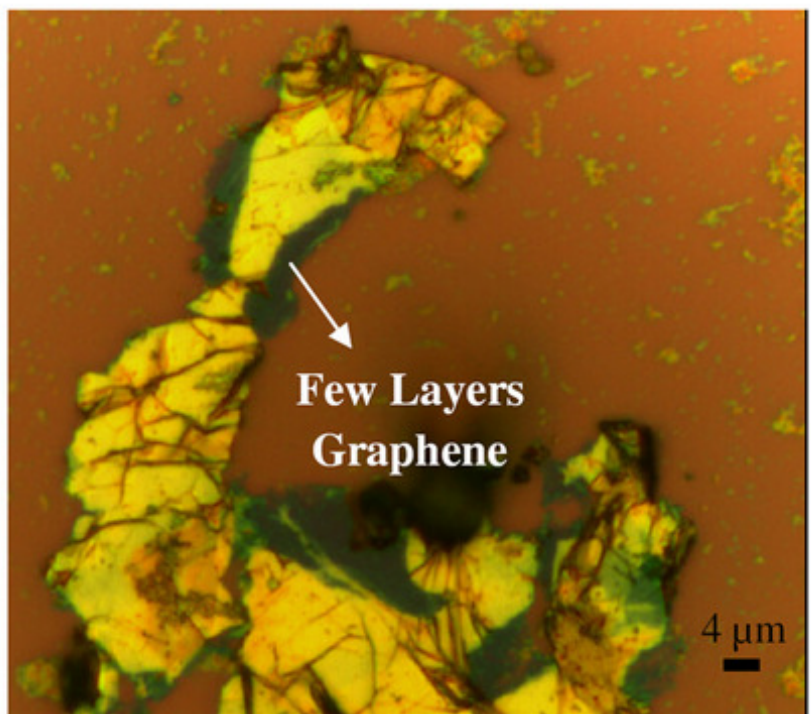

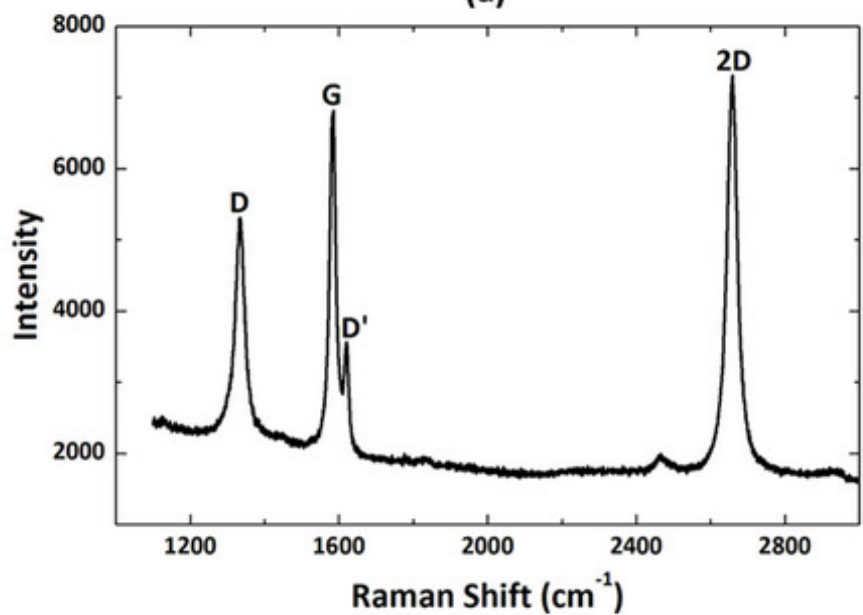

**Fig.7.** Optical Microcopy of Few layer Graphene formed on top of copper and then transferred to Si/SiO$_2$ (a) and its Raman spectrum (b)

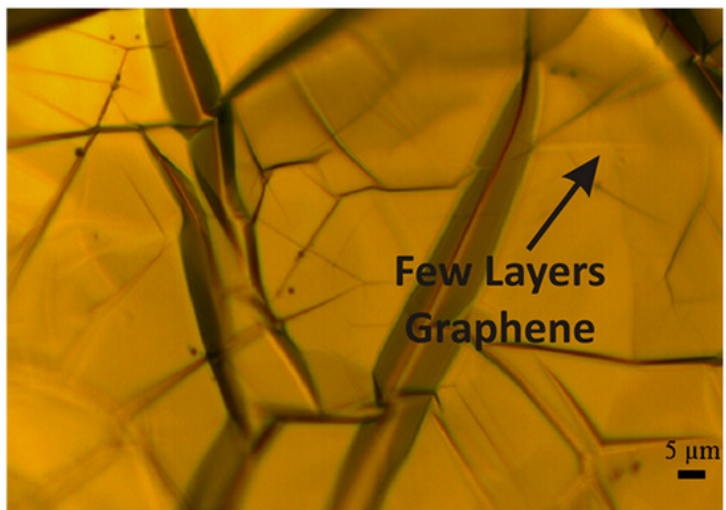

(a)

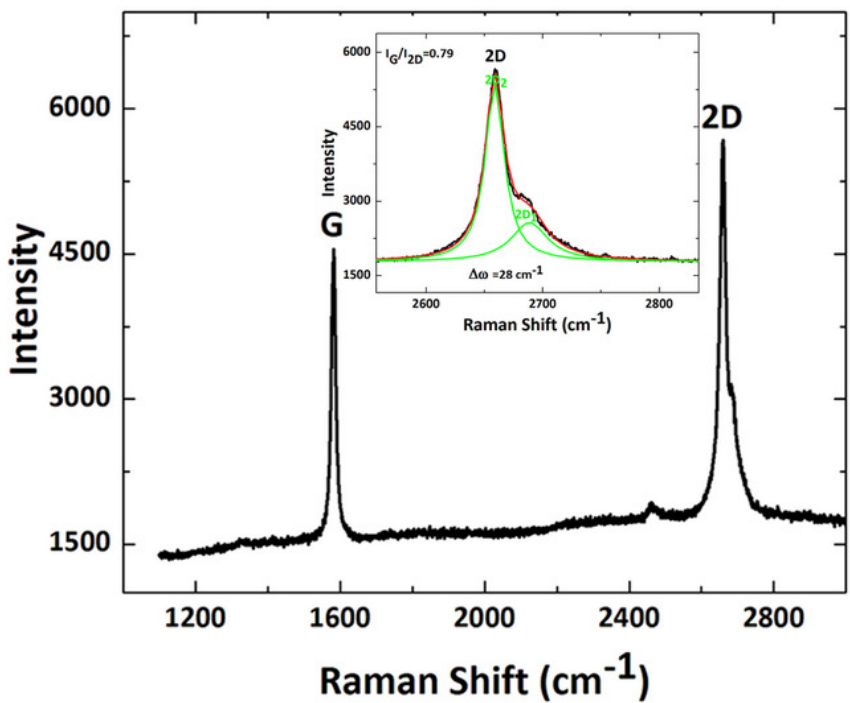

(b)

**Fig.8.** Optical Microcopy of Few layer Graphene formed on top of nickel (a) and Raman spectrum of the formed layer and its 2D band deconvolution (b)

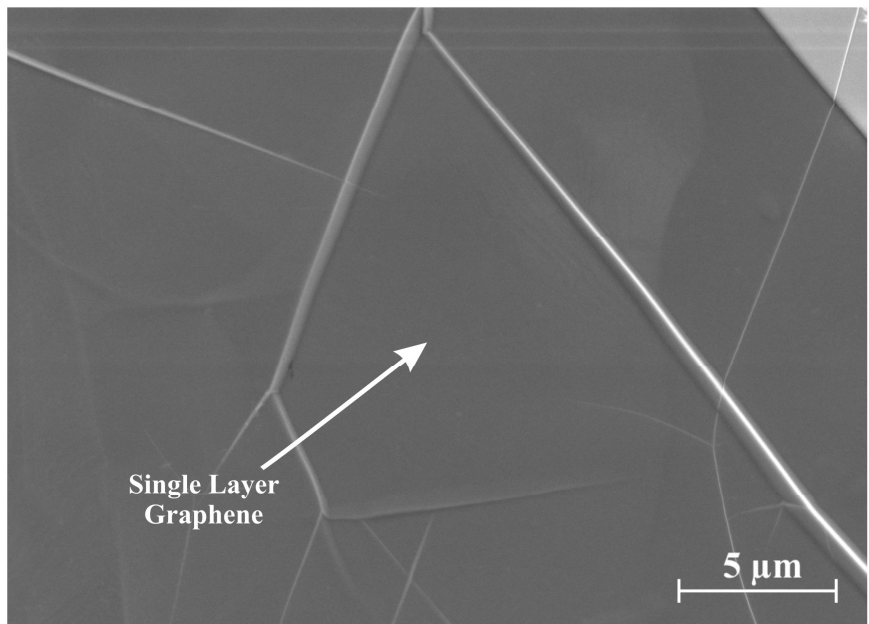

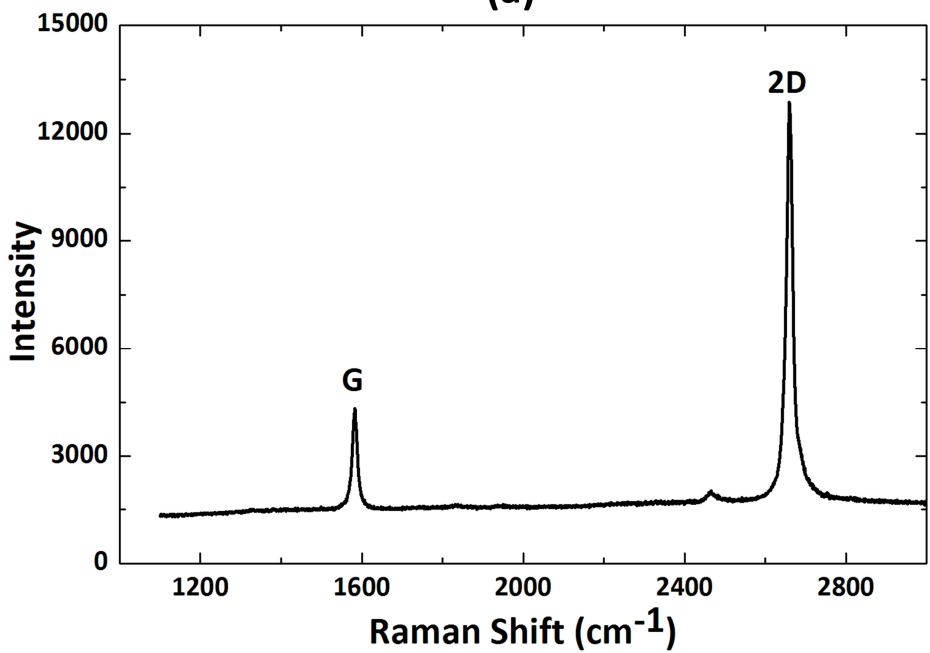

**Fig.9.** SEM photo of Single layer graphene formed on nickel (a) and its Raman Spectrum (b)